\documentclass[utf8]{arxiv}   %

\usepackage{url,hyperref,lineno,microtype,subcaption}
\usepackage[onehalfspacing]{setspace}
\usepackage{multirow}
\usepackage{graphicx}
\usepackage{pgf}
\usepackage{amsmath}
\usepackage{comment}
\usepackage{pifont}
\usepackage{array}
\usepackage{tabularx}
\usepackage{diagbox}
\usepackage{booktabs}
\usepackage{arydshln}

\newcommand{\PreserveBackslash}[1]{\let\temp=\\#1\let\\=\temp}
\let\oldding\ding
\renewcommand{\ding}[2][1]{\scalebox{#1}{\oldding{#2}}}
\newcommand{\xmark}{\textcolor{red}{\ding[2]{55}}}%
\newcolumntype{M}[1]{>{\centering\arraybackslash}m{#1}}

\def\keyFont{\fontsize{8}{11}\helveticabold }
\def\firstAuthorLast{Mahmood et al.} 
\def\Authors{Usman Mahmood\hspace{1pt}$^{1,*}$, Robik Shrestha\hspace{1pt}$^{2}$, David D.B. Bates\hspace{1pt}$^{3}$, Lorenzo Mannelli\hspace{1pt}$^{4}$, Giuseppe Corrias\hspace{1pt}$^{5}$, Yusuf Erdi\hspace{1pt}$^{1}$ and Christopher Kanan\hspace{1pt}$^{2}$}

\def\Address{$^{1}$ Department of Medical Physics, Memorial Sloan Kettering Cancer Center, New York, New York, USA \\
$^{2}$Chester F. Carlson Center for Imaging Science, Rochester Institute of Technology, Rochester, New York, USA  \\ 
$^{3}$Department of Radiology, Memorial Sloan Kettering Cancer Center, New York, New York, USA  \\
$^{4}$Institute of Research and Medical Care (IRCCS) SDN, Naples, Italy \\
$^{5}$Department of Radiology, University of Cagliari, Cagliari, Italy}

\def\corrAuthor{Usman Mahmood}
\def\corrEmail{mahmoodu@mskcc.org}

\begin{document}

\onecolumn
\firstpage{1}

\title[]{Detecting Spurious Correlations with Sanity Tests for Artificial Intelligence Guided Radiology Systems} 

\author[\firstAuthorLast ]{\Authors} 
\address{} 
\correspondance{} 
\extraAuth{}
\maketitle
\begin{abstract}
Artificial intelligence (AI) has been successful at solving numerous problems in machine perception. In radiology, AI systems are rapidly evolving and show progress in guiding treatment decisions, diagnosing, localizing disease on medical images, and improving radiologists' efficiency. A critical component to deploying AI in radiology is to gain confidence in a developed system's efficacy and safety. The current gold standard approach is to conduct an analytical validation of performance on a generalization dataset from one or more institutions, followed by a clinical validation study of the system's efficacy during deployment. Clinical validation studies are time-consuming, and best practices dictate limited re-use of analytical validation data, so it is ideal to know ahead of time if a system is likely to fail analytical or clinical validation. In this paper, we describe a series of sanity tests to identify when a system performs well on development data for the wrong reasons. We illustrate the sanity tests' value by designing a deep learning system to classify pancreatic cancer seen in computed tomography scans.
\tiny
 \keyFont{ \section{Keywords:} Deep learning, Computed Tomography, Bias, Validation, Artificial Intelligence, Spurious Correlations} 
\end{abstract}

\section{Introduction}

Artificially intelligent (AI) computer-aided diagnostic (CAD) systems have the potential to help radiologists on a multitude of tasks, ranging from tumor classification to improved image reconstruction~\citep{jin2020artificial,el2020artificial, yala2019deep, antonelli2019machine}. To deploy medical AI systems, it is essential to validate their performance correctly and to understand their weaknesses before being used on patients~\citep{laghi2020cautions,rudin2019stop,the2018digital}. For AI-based software as a medical device, the gold standard for analytical validation is to assess performance on previously unseen independent datasets~\citep{bluemke2020assessing,soffer2019convolutional,kim2019design,el2018machine}, followed by a clinical validation study. Both steps pose challenges for medical AI. First, it is challenging to collect large cohorts of high-quality and diverse medical imaging data sets that are acquired in a consistent manner \citep{recht2020integrating, parmar2018data}. Second, both steps are time-consuming, and best practices dictate limited re-use of analytical validation data. The cost of failing the validation process could prohibit further development of particular applications. 

One reason AI systems fail to generalize is that they learn to infer spurious correlations or covariates that can reliably form decision rules that perform well on standard benchmarks~\citep{geirhos2020shortcut}. For example, an AI system successfully trained to detect pneumonia from 2D Chest X-rays gathered from multiple institutions, but it failed to generalize when images from new hospitals outside of the training and assessment set were used to evaluate the system~\citep{zech2018variable}. The investigators found that the system had unexpectedly learned to identify metal tokens seen on the training and assessment images~\citep{zech2018variable}. In hindsight, the tokens were obvious spurious correlators, but in other cases, the covariates can be less obvious~\citep{geirhos2020shortcut}. For example, subtle image characteristics that may be unrelated to the target object, such as high-frequency patterns~\citep{jo2017measuring,kafle2019challenges,ilyas2019adversarial}, object texture~\citep{geirhos2019imagenet,baker2018deep}, or intangible attributes of objects are known to cause AI systems to form decision rules that may not generalize~\citep{sinz2019engineering, geirhos2020shortcut}. Current research has focused on explaining or interpreting AI decisions using various visualization techniques~\citep{reyes2020interpretability}, but these do not necessarily imply that a system will generalize~\citep{adebayo2018sanity,kim2018interpretability,ghorbani2019interpretation, lakkaraju2020fool}. 

Addressing system failures before clinical deployment is critical to ensure that medical AI applications are safe and effective. Identifying systems that are right for the wrong reasons during the development stages can expedite development by not wasting valuable validation data from multiple institutions or conducting doomed clinical validation studies.

The standard approach used to identify system failures involves testing with held-out development or generalization test datasets \citep{park2018methodologic}. However, development test sets are subsets of the training data, and their primary value lies in identifying systematic errors or bugs within the AI algorithm. Generalization test data are independent of the development data (i.e., their joint probability distribution of inputs and labels differ from training and development test data) \citep{teney2020value}. The generalization data's value is to assess how well a trained model may adapt to previously unseen data. However, neither type of test is sufficiently robust enough to declare when an AI system is ready for the clinic. 

We provide a set of sanity tests that can demonstrate if a trained system is right for the wrong reasons. We developed a weakly supervised deep learning system for classifying pancreatic cancer from clinical computed tomography (CT) scans to illustrate their use. Our main contributions are:

\begin{enumerate}
    \item We provide a set of sanity tests to determine if a system is making predictions using spurious correlations in the data. 
    \item We describe a system for using deep learning with CT images to detect pancreatic cancer, and we apply our set of sanity tests to both development and generalization test datasets. We train and assess four unique variants of this system to illustrate the pipeline and demonstrate that the system looks as if it performs well in many scenarios, but it is predicting using spurious correlations.
    \item We illustrate how to use a method to generate noise images from the patients' volumetric CT scans. These can then be used to assess the influence of noise on the AI system's performance. 
\end{enumerate}

\section{MATERIALS AND METHODS}

\subsection{Sanity Tests for AI Systems}
There are various testing procedures employed in software engineering to determine if a system is working correctly, such as smoke and sanity tests~\citep{gupta2013}. Smoke tests evaluate the critical functionality of a system before conducting additional tests. In AI, this is analogous to reaching an acceptable level of performance on the development test data, which matches the training data's distribution. Development test data is typically a random sample of the training data (e.g., 30\% test and 70\% train). The stopping point for many AI projects is when acceptable performance is achieved on the development test set, but in software engineering, the next step is to conduct `sanity tests' that indicate if a system produces obvious false results. If the sanity tests fail, further development is done before conducting more time-consuming and rigorous tests, which for AI systems used in medical applications could correspond to analytical and clinical validation studies. For AI systems, sanity tests would identify if a system is achieving good results on the development test set for the wrong reasons (e.g., covariates or spurious correlations) and will therefore fail in other environments or on other datasets. 

Sanity tests are occasionally used to identify if a system is unlikely to generalize \citep{adebayo2018sanity, winkler2019association, oakden2020hidden}. However, the tests are often designed to evaluate literature methods instead of being used as a crucial development tool. For example, Shamir et al. critiques the methods by which face datasets were designed and evaluated by showing that commonly used face recognition datasets were classified correctly even when no face, hair, or clothing features appeared in the training and testing datasets~\citep{shamir2008evaluation}. As another example, in response to a report suggesting AI systems could diagnose skin cancer at the level of dermatologists \citep{esteva2017dermatologist}, Winkler et al. evaluated the limits of the claim by testing a trained AI system using dermoscopic images where the covariate's, hand-drawn skin markings, were first present and then absent from pictures of the skin cancer. They observed that when skin markings were present, the probability that the AI system classified images as having skin cancer increased significantly. With the markings removed, the probability decreased, which led them to conclude that the AI system associated the markings with cancer instead of the actual pathology \citep{winkler2019association}.

For AI-based medical devices, conducting sanity tests can prevent needless harm to the patient and save a considerable resources. However, without sufficiently large, well-annotated datasets, performing analytical validation to determine the root causes that drive AI systems to fail before deployment remains a challenge~\citep{willemink2020preparing}. Moreover, after independent testing data is gathered, regulatory organizations advise that the data be used a limited number of times to prevent over-fitting ~\citep{petrick2013evaluation}. For example, the United States Food and Drug Administration ``discourages repeated use of test data in the evaluation'' of CAD systems~\citep{us2020clinical}. Clinical validation of deployed systems is likewise time-consuming to organize and often costly. 

\begin{figure}
	\centering
			\includegraphics[width=\textwidth]{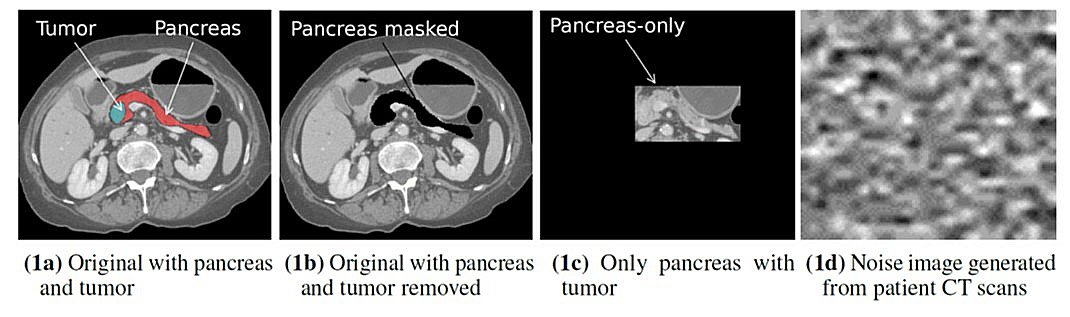}
	\caption{A sample cross sectional axial slice from a patient CT scan processed into the different input formats that were used for the sanity tests. The AI system is trained using each of these input format scenarios and then tested on all of them. \textbf{a)} The original image with the pancreas and tumor present. \textbf{b)} The original image with the pancreas and tumor removed. \textbf{c)} The anatomy surrounding the pancreas is cropped and only the pancreas and tumor remain. \textbf{d)} Noise image generated from the patient CT scans. Although this figure shows a single slice, the same processing is applied to all classes, slices and patients in the datasets. A system that is trained to identify the target organ or disease should achieve chance performance when tested with scans where the target or disease is removed. \label{fig:sanityEX} } 
\end{figure}

We propose a series of sanity tests to identify if an AI system may fail during the development phases and before conducting more extensive generalization tests. We also describe how the tests are used with a case study to detect pancreatic cancer from weakly labeled CT scans. The tests are as follows:

\begin{itemize}
    \item \textbf{Train and test with the target-present and absent}. If an AI system is trained to distinguish between normal and abnormal diagnostic features (e.g., organ with cancer shown in Fig.~\ref{fig:sanityEX}a), then it should fail when that target is removed from the development test data (e.g., Fig.~\ref{fig:sanityEX}b). If the system still works effectively after removing the target from testing data, then that indicates it is confounded. In our case study, this corresponds to removing the pancreas from normal scans and pancreas with tumor from abnormal scans using a segmentation mask, as shown in Fig.~\ref{fig:sanityEX}c. We removed the whole pancreas because the pancreatic tumor often distorts the contours of the surrounding anatomy \citep{galvin2011part}. 
    
     \item \textbf{Train and test the system with background patches or noise images}. Background patches consist of non-target regions of the image. Noise images can be generated from the volumetric CT scans in the development and generlization datasets. Both can determine if the different classes can be discriminated based on features unrelated to the target objects \citep{shamir2008evaluation}. If classes are discriminated against with high confidence using the noise image types, then the system is confounded, and it is using features of the image acquisition process to delineate classes. An example noise image generated from the patient CT scans is shown in Fig.~\ref{fig:sanityEX}d.  
     
    \item \textbf{Test with different regions of interests (ROIs)}. Training and testing AI systems on precisely outlined segments of images does not reflect real-world usage. Medical centers, private practices, or institutes where AI is deployed may not have the resources or expertise to precisely outline the anatomical area~\citep{price2019medical}.Furthermore, similar to radiologists, AI systems may have to parse through anatomy they have never encountered during training. Therefore, it is desirable to ensure that when systems are trained on a select portion of images, as shown in Fig. \ref{fig:sanityEX}c, they can generalize to the original image shown in Fig. \ref{fig:sanityEX}a. 
\end{itemize}
These sanity tests can be conducted solely using the development dataset, but ideally they would also be used in conjunction with another generalization dataset. They require four input formats, as shown in Fig. \ref{fig:sanityEX}, to be generated from the same development dataset. 

\begin{table}[ht]
\caption{Scan parameters and patient-specific characteristics for development and generalization data.}
\resizebox{\textwidth}{!}{%
\begin{tabular}{rccc}
\hline
\multicolumn{1}{l}{} & \multicolumn{2}{c}{Development Data: Train, Tune, \& Test} & \multirow{2}{*}{Generalization Data} \\
\multicolumn{1}{l}{} & TCIA - Pancreas CT & \begin{tabular}[c]{@{}c@{}}Medical Image Segmentation \\ Decathalon (MSD)\end{tabular} &  \\ \hline
Annotated & Yes & Yes & No \\
CT Vendor & Phillips and Siemens & General Electric & General Electric \\
CT Model & ** & LS16 or HD750 & HD750 \\
Total \# of Patients & 82 (27 female/55 male) & 281* & 116 (61 female/55 male) \\
\# used to train & 58 & 60 & NA \\
\# used to tune & 15 & 14 & NA \\
\# used to test & 9 & 8 & 116 \\
\multicolumn{1}{c}{\textbf{Dataset Information:}} &  &  &  \\
Average age (min to max) & 46.8 (18 to 76) & ** & 63 ( 18 to 90) \\
Scan start time after contrast administration & $\sim$70s & 80 to 85s & $\sim$40s \\
Avg. \# of total slices (min/max) & 256 (181 to 466) & 95 (37 to 751) & 186 (102 to 278) \\
Avg. \# of slices consisting of only pancreas (min/max) & 85 (45 to 144) & 30 (11 to 147) & NA \\
 &  &  &  \\
\multicolumn{1}{c}{\textbf{Scan parameters:}} & \multicolumn{1}{l}{} & \multicolumn{1}{l}{} &  \\
Tube potential (kVp) & 120 & 120 & 70 keV (80/140 kVp) \\
Slice thickness (mm) & 1.5 - 2.5 & 2.5 & 2.5 \\
Pixel dimensions (mm) & 0.664 to 0.977 & 0.606 to 0.977 & 0.547 to 0.976 \\
Tube current modulation index & ** & \begin{tabular}[c]{@{}c@{}}Noise Index: \\ 14 (HD750) / 12.5 (LS16)\end{tabular} & NA \\
Tube current (mA) min to max range & ** & 220–380 mA & 260-600 \\
Rotation time (s) & ** & 0.7 (HD750) / 0.8 (LS16) & 0.7 (HD750) \\
Pitch & ** & 0.984 (HD750) / 1.375 (LS16) & 0.984 (HD750) \\
Reconstruction algorithm & ** & ** & ***FBP/ASiR 20\% \\
Reconstruction kernel & ** & ** & Standard \\
Iterative reconstruction strength & ** & ** & 20\% \\
\# of data channels & ** & ** & 64 \\
Size of a single data channel (mm) & ** & ** & 0.625 \\
Bowtie filter & ** & ** & Large Body \\
CT scan series released or used & Axial portal venous phase & Axial portal venous phase & Axial arterial phase \\ \hline
\end{tabular}%
}
\footnotesize{$^*$A subset of the MSD dataset was randomly selected to train the model \\
$^{**}$Not available in accompanied report or DICOM header \\
$^{***}$FBP - Filtered Back Projection,  ASiR - Adaptive Statistical Iterative Reconstruction \\
$^{****}$LS16: LightSpeed16, HD750: Discovery High Definition 750}
\label{tab:patientChar}
\end{table}

\subsection{Datasets}

Institutional review board approval was obtained for this Health Insurance Portability and Accountability Act-compliant retrospective study. The requirement for informed consent was waived. The case study is designed as a binary classification problem with the aim of identifying patients who have pancreatic cancer versus those who do not from the provided CT scans. We distinguish between the development dataset used for training, tuning and testing, and the generalization test data used to validate the efficacy of the system. 

\textbf{Development Data.} The development dataset consisted of patient CT scans collected from two open-access repositories where detailed annotations were available. The normal pancreas CT scans were obtained from The Cancer Imaging Archive Normal (TCIA) Pancreas Dataset with 82 contrast-enhanced abdominal CT scans ~\citep{roth2016data}. Seventeen patients from the TCIA dataset were reported to be healthy kidney donors. The remaining patients were selected because they had no major abdominal pathology or pancreatic lesions~\citep{roth2016data}. The abnormal pancreas CT scans were obtained from the  Medical Image Segmentation Decathlon (MSD) dataset, consisting of abdominal CT scans from 281 patients. The MSD dataset contains patients who presented with intraductal mucinous neoplasms, pancreatic neuroendocrine tumors, or pancreatic ducal adenocarcinoma~\citep{simpson2019large}. They were originally used to predict disease-free survival or assess high-risk intraductal papillary mucinous neoplasms seen on the CT scans \citep{simpson2019large}. We randomly selected 82 cases from the MSD dataset to match the TCIA dataset size to avoid class-imbalance issues. The development data were randomly split into a training (58 normal, 60 cancer), tuning (15 normal, 14 cancer), and held-out test (9 normal, 8 cancer) set. To ensure the number of positive and negative samples were balanced in each split, we used stratified 5-fold cross-validation for training. Table~\ref{tab:patientChar} shows the patient demographics and scanning parameters provided for each dataset. 

\textbf{Generalization Data: Dual Energy CT (DECT).} The generalization data consists of 116 patients (58 without PC, 58 with PC) who received routine DECT scans between June 2015 to December 2017 (see Table~\ref{tab:patientChar}). The patients without pancreatic cancer received DECT CT Urography (CTU) exams and were selected based on the statement of a negative or unremarkable pancreas and liver in the original radiologist report. Those with cancer were selected if they had undergone a DECT arterial phase CT scan and were histologically confirmed to have pancreatic cancer. All patients were scanned on a 64 slice CT scanner (Discovery CT750 HD, GE Healthcare, Milwaukee, WI, U.S.) with rapid switching DECT following the administration of 150 mL of iodinated contrast (Iohexol 300 mgI/mL, Omnipaque 300, GE Healthcare, Cork, Ireland), at 4.0 mL/s. The scan parameters are displayed in Table~\ref{tab:patientChar}. With DECT, multiple image types can be generated, such as virtual monochromatic images (VMI) that depict the anatomy and physiology from the viewpoint of a monochromatic x-ray source \citep{hsieh2003computed}. The VMI scans can be reconstructed at energies ranging from 40 to 140 keV. For this study, all scans were reconstructed at 70 keV because of its use in the clinic. The images were generated using the GSI MD Analysis software available on Advantage Workstation Volume Share 7 (GE Healthcare). Those patients who had a history of surgery and liver abnormalities were excluded from the test set, as were any patients who had metal adjacent to the pancreas or visible artifacts on the scans. This dataset was not used during the training or tuning stages. 

\subsection{AI System - CTNet}

\begin{figure}[t]
\begin{center}
\includegraphics[width=1.\linewidth]{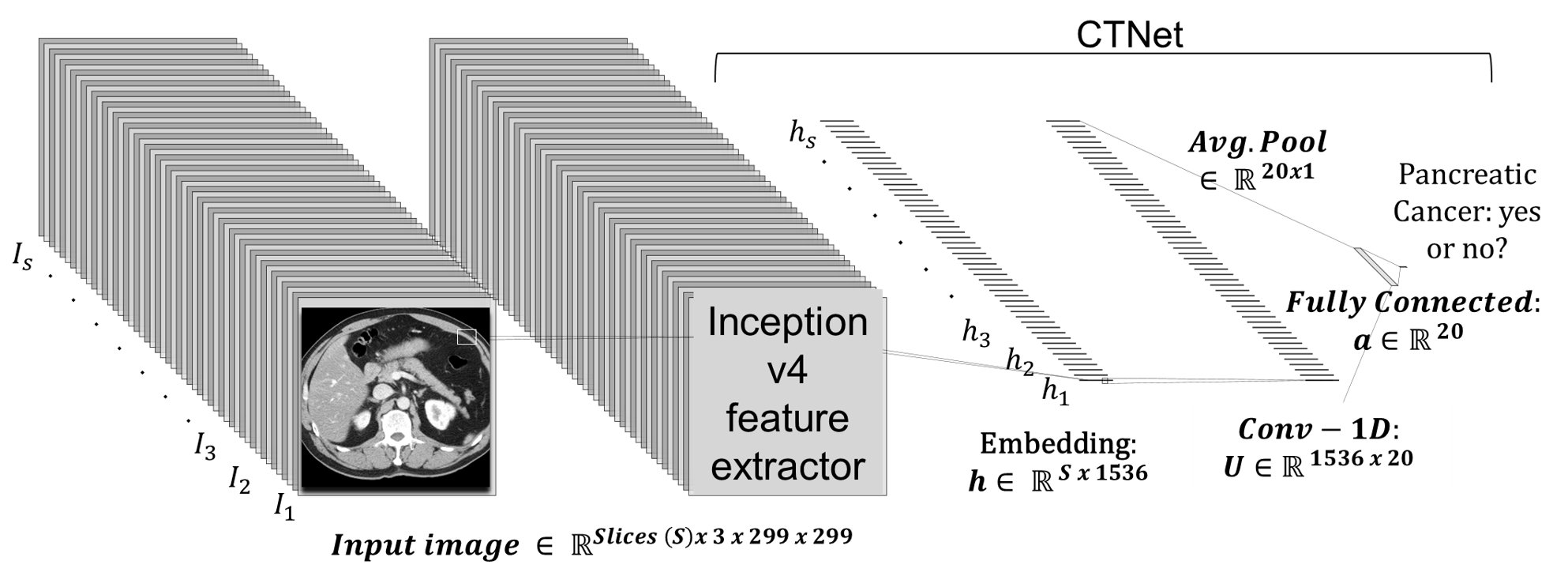}
\end{center}
\caption{\textbf{CTNet architecture.} CTNet takes as input a volumetric CT scan and outputs a classification prediction. Features are extracted from each slice of the CT scan by the Inception v4 network. The output feature vector is then reduced in dimension with a single convolutional layer, followed by an adaptive average pooling operation applied over the number slices. The resulting vector is fed into a fully connected layer, which has a single output.}\label{fig:CTNet}
\end{figure}

The prediction system is dubbed CTNet. It is designed to map a 3D CT scan to a probability estimate that indicates if pancreatic cancer is present or not. CTNet closely resembles systems in literature that use ImageNet pre-trained convolutional neural networks (CNNs) on radiology scans~\citep{draelos2020machine, raghu2019transfusion, winkler2019association, bien2018deep, esteva2017dermatologist, liu2017detection, paul2016deep}. The model architecture is shown in Fig.~\ref{fig:CTNet}.

Given a total set of $s$ slices in a scan, where each individual slice $t$ is a $299 \times 299$ image, an ImageNet pre-trained Inception v4 CNN was used to extract an embedding $\mathbf{h}_t \in \mathbb{R}^d$ from each slice. The embeddings were extracted from the penulmitate layer, which renders a $d=1536$ dimensional feature vector for each image~\citep{szegedy2016inception}. Because Inception v4 is designed to take as input a $299 \times 299 \times 3$ RGB image, we replicated each slice to create faux RGB images. Following others~\citep{van2015off}, the CNN was not fine-tuned for CT data.

After extracting the embeddings from all scan slices, they are fed into a neural network that makes the final prediction, which is given by
\begin{equation}
P\left( {Cancer = 1|{\mathbf{h}}_1 ,{\mathbf{h}}_2 , \ldots {\mathbf{h}}_s } \right) = \sigma \left( {b + \frac{1}
{s}{\mathbf{w}}^T \sum\limits_{t = 1}^s {\operatorname{ReLU} \left( {{\mathbf{Uh}}_t  + {\mathbf{a}}} \right)} } \right),
\end{equation}
where $\sigma \left( \cdot \right)$ denotes the logistic sigmoid activation function, $b \in \mathbb{R}$ is the output layer bias, $\mathbf{w} \in \mathbb{R}^{20}$ is the output layer's weight vector, $\mathbf{U} \in \mathbb{R}^{{20} \times 1536}$ is the hidden layer weight matrix, $\mathbf{a} \in \mathbb{R}^{20}$ is the hidden layer bias, and ReLU is the rectified linear unit activation function. In preliminary studies, we found that using 20 hidden units sufficed to achieve strong performance. If the predicted probability was above 0.5, then the patient was classified as positive for pancreatic cancer. 

The model was trained using the binary cross-entropy loss function with a mini-batch size of 1. The weights were initialized using the Kaiming method. For all systems trained in this study, we used the Adam optimizer with~\citep{kingma2014adam} a base learning rate of $1e^{-4}$, $L_2$ weight decay of $1e^{-6}$, and bias correction terms, $\beta_{1} = 0.9$ and $\beta_{2} = 0.999$. The learning rate was reduced by a factor of 2 over the course of training when the validation loss had stopped improving. The system was trained for 100 epochs. Since our training dataset was balanced, we did not scale the loss for any particular class's prevalence. During training, no data augmentation techniques were applied. The model was implemented in Python 3.8 with PyTorch 1.6.0 on a computer with a 12 GB NVIDIA Titan V GPU.

\subsection{Scan Preprocessing}

Since the voxel size varied from patient to patient, the CT scans were first resampled to an isotropic resolution of 1.0 $\times$ 1.0 $\times$ 1.0 mm using SINC interpolation. They were then resized to a height and width of 299 $\times$ 299 pixels using bilinear interpolation, which is the original input image size used to train the Inception v4 network. The voxel Hounsfield unit (HU) value was clipped to be between $\pm 300 HU$ and normalized to have zero mean and unit variance (i.e., $[0,1]$). Normalization was performed by subtracting the mean and dividing by the standard deviation computed from the training dataset. This processing was applied to both the development and generalization datasets. 

\subsection{Noise Image Generation}
\begin{figure}[t]
	\centering
		\includegraphics[width=\textwidth]{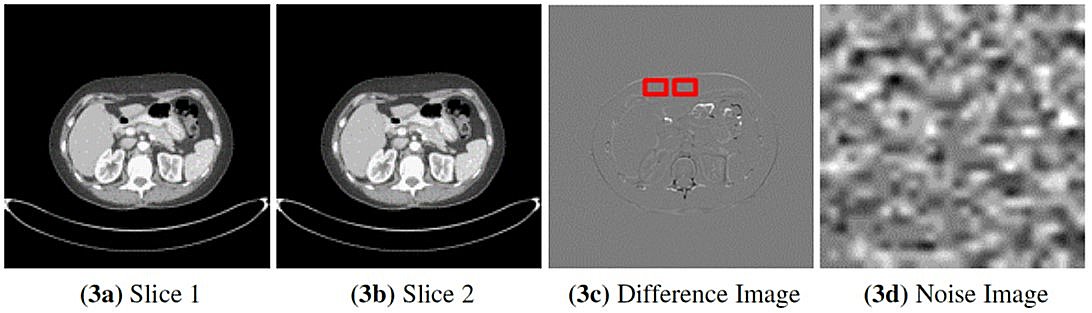}
	\caption{Method to obtain noise maps from sequentially acquired images. Two sequential slice images are subtracted from each other (\textbf{a} and \textbf{b}). \textbf{c)} Difference image resulting from subtraction showing the sliding 30 $\times$ 30 pixel window used to extract uniform patches. \textbf{d)} All patches were averaged to generate a single noise image. \label{fig:noise}} 
	\end{figure}

The noise images are used to determine if the institutional scanning practices or noise characteristics of the imaging systems confound the classification results. A key characteristic of the image is that they must be uniform and devoid of any perceptible patterns. We generated noise images from each patient's CT scan using an approach similar to \citep{christianson2015automated,tian2016accurate}, and as shown in Fig. 4. For a scan with $s$ sequential slices, where each slice $t$ is an image $I_t \in \mathbb{R}^{299\times299}$, we subtract adjacent slices to produce $s-1$ difference images $D_I$, where $D_I = I_t - I_{t-1}$ and $1 \le t < s$~. The subtraction process eliminates most of the anatomical features seen in the scan. We then apply a Sobel edge enhancing filter to each $D_I$ to identify and remove any remaining anatomical patterns. Then we loop through each $D_I$ to extract non-overlapping patches of size $30 \times 30$ pixels. The patch size was selected to minimize the impact of the non-uniformity of the CT HU values within the region of interest (e.g., due to streaking or beam hardening artifacts)~\citep{tian2016accurate}. However, patches of transitional boundary areas (i.e., interface between different tissue types) consisted of discernible patterns that could be spuriously correlated with the class labels. Consequently, to identify and exclude boundary patches, we generated and analyzed each patch's histogram. First-order statistical measures, such as skewness, kurtosis, and the standard deviation and the number of peaks within the histogram were used to identify and exclude boundary patches. Histograms with a skewness value within $\pm 0.1$, kurtosis of $3.0 \pm 0.5$, a standard deviation less than 16, and those with a single peak were included. Published descriptions of the noise image generation method do not provide choices for each of the parameters, so we chose them via visual inspection to eliminate transition areas or edges. The patches that met the criteria were then averaged together to create a single noise image representation of size $30 \times 30$ for the $D_I$, as shown in Fig. \ref{fig:noise}. Finally, the $s-1$ noise images for the patient were upsampled using SINC interpolation to a dimension of $299 \times 299$. 

\subsection{Experiments}
\label{sec:experiments}
To employ the sanity tests, we processed four representations or input formats of the training, tuning, and held out development test sets. Representative images are shown in Fig.~\ref{fig:sanityEX}. For the first format, termed pancreas-only (Fig.~\ref{fig:sanityEX}c), we used the provided annotations to crop out the area surrounding the normal pancreas and the pancreas with tumor from each scan so that only the target areas remained. The second format, shown in Fig.~\ref{fig:sanityEX}a, consisted of the original uncropped scans. For the third format, the annotations were used to remove the pancreas and pancreas with tumor from the original uncropped scans, Fig.~\ref{fig:sanityEX}b. The fourth format consisted of the noise images. We trained four systems, one for each input format, and tested each of them with the held-out test sets of the other formats. Since annotations were not available for the generalization test set, we generated two formats: 1) the original uncropped images and 2) the noise images. We performed stratified 5-fold cross-validation with the same division of scans across the four systems. For this study, we consider the baseline against which all results are compared to be the system trained with the pancreas-only scans shown in Fig.~\ref{fig:sanityEX}c, as it should be the representation that maximizes the signal to noise ratio. 

\subsection{Statistical Analysis}

Each system's classification performance was assessed using the area under the receiver operator characteristic curve (AUC). We report the average AUC and corresponding 95\% confidence interval (CI) across cross-validation runs. An average AUC score of 1.0 represents perfect classification performance. The average AUC across runs and the corresponding confidence intervals were determined using R (Rstudio version 3.6.2) with the package cvAUC for cross-validated AUC~\citep{ledell2015computationally}. In addition to confidence intervals, statistically significant differences between test runs was confirmed with the DeLong test statistic for AUCs~\citep{delong1988comparing}. The level of significance was set at $P \le .05$.  Table~\ref{tab:sanity-tests} provides an overview of how the sanity tests should be interpreted and implemented in practice.

\begin{table}[ht]
\centering
\caption{The proposed sanity tests to assess the reliability of medical AI systems. \label{tab:sanity-tests}}
\begin{tabular}{p{0.4\textwidth}p{0.35\textwidth}M{0.15\textwidth}}
\hline
\textbf{Sanity Test} & \textbf{Implications of Failing the Test} & \textbf{Does CTNet Pass the Test?} \\
\hline

\textbf{Train \& test with and without the target:} The system should achieve an AUC of around 0.5 when tested without the target in test images. & Images contain spurious covariates  that can be exploited by the model. & \vspace{6mm} \xmark \\ \\

\textbf{Train \& test using noise images:} The system should achieve an AUC of around 0.5 on test data. & Classification performance cannot be attributed to recognition of the target (i.e., covariates contribute to the learned classification decision rule). & \vspace{6mm} \xmark \\ \\

\textbf{Test system with different sized ROIs:} The additional or reduced context should not alter the performance. & The system cannot decorrelate features of the target from its co-occuring context (i.e., Contextual Bias~\citep{singh2020don}). & \vspace{6mm} \xmark \\ \\
\hline
\end{tabular}
\end{table}

\section{Results}

\begin{figure}[t]
    \centering
        \includegraphics[width=1.0\textwidth]{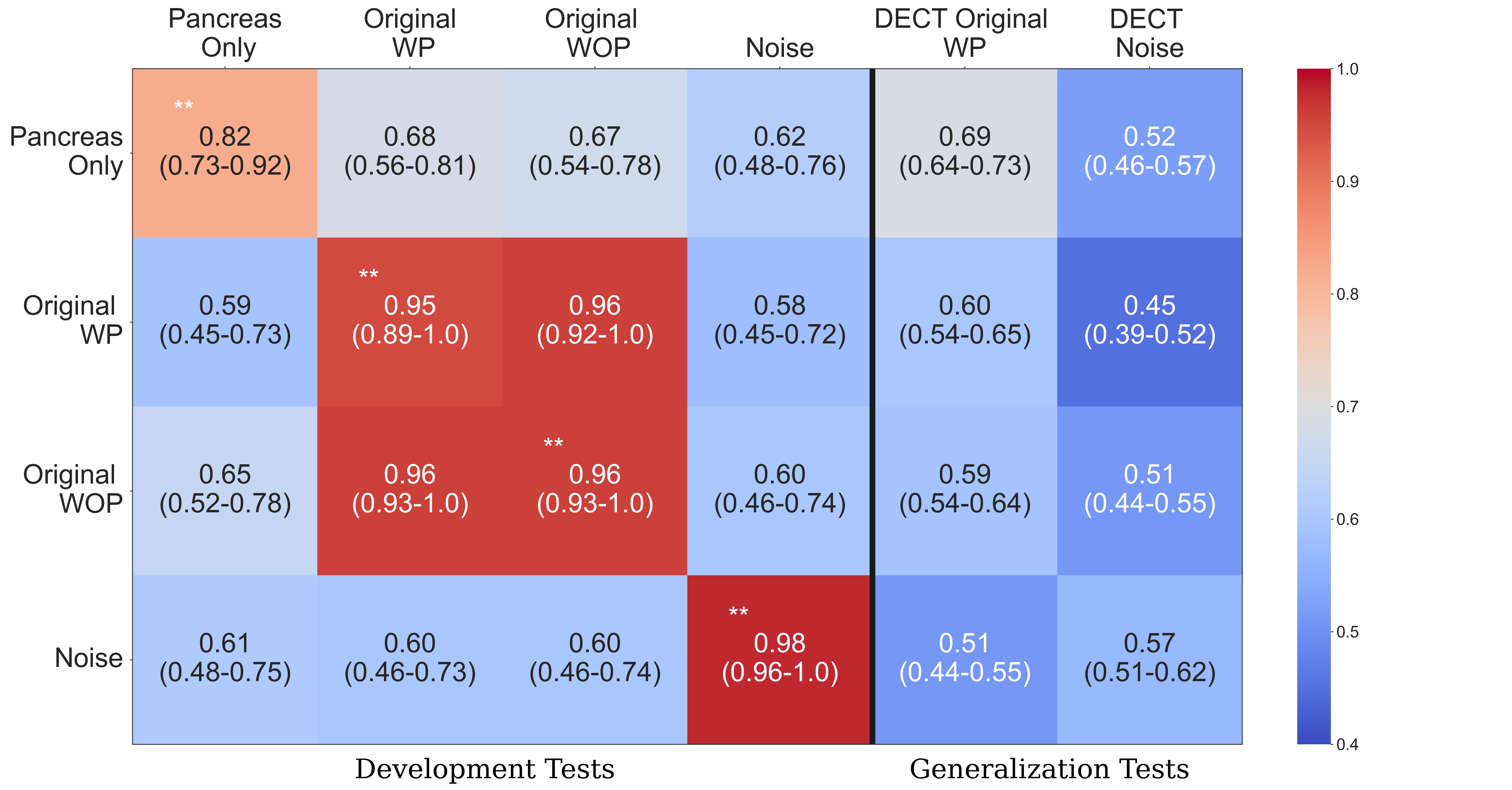}
    \caption{\textbf{AUC heatmap across models for each input format type.} Each row indicates the cross-validated mean AUC with 95\% confidence intervals for the systems trained with a given input format and assessed across all input format variants on the development dataset (first four columns) and generalization dataset (last two columns). The diagonal elements on the development tests correspond to training and testing with the same input format. The last two columns show mean AUC on the generalization dataset with the original and the noise formats. Red indicates the highest AUC values, while light blue indicates the lowest AUC values. The results indicate that the system correctly classifies patients using spurious correlations, instead of features specific to pancreatic cancer. **Test images processed identically to the data used for training that model. DECT = Dual Energy CT.}
    \label{fig:main-results}
\end{figure}

Our main results are shown in Fig.~\ref{fig:main-results}. Each row corresponds to the cross-validated AUC obtained when training on one input format and testing on all other formats. The diagonal elements for the development tests correspond to training and testing on the same input format  (i.e., self-tests), while the others represent AUC scores from training on one variant and testing on the another (i.e., non-self tests). We expect a system trained on one format to perform the best on test data processed in an identical manner, which is consistent with our results. The system trained with the pancreas-only images achieved an AUC of 0.82 (95\% CI: 0.73 - 0.92) on its self-test set. If the system was considered to pass the sanity tests, we would expect it to have the highest AUC across all input formats. However, instead, it is the lowest among the self-tests. Its performance is significantly lower ($\textit{P}<.001$) than systems trained on the original scans, with and without the pancreas present, 0.95 AUC (95\% CI: 0.89 - 1.0) and 0.97 AUC (95\% CI: 0.93 - 1.0), respectively. The noise only system achieved the highest AUC of 0.98 (95\% CI: 0.96 - 1.0) on its self test set. 

Our results indicate that this dataset with CTNet fails all of the sanity tests. The self-tests for the noise images and the no pancreas format achieved the highest AUCs rather than being near a score of 0.5. This result on the development data is confirmed on the generalization tests where these system's performances are significantly lower lower than on the development data self-tests ($\textit{P}<.001$). 

\section{Discussion}

Identifying covariates that cause unintended generalization or those that cause machines to fail unexpectedly in deployment remains a challenge across deep learning applications. We described sanity tests that could reveal if covariates drive classification decision-making and tested them with a case study designed to classify pancreatic cancer from CT scans. Failing these sanity tests provides an early indicator of potential biases being responsible for the observed performance and that further in the development process, a system will unintentionally generalize or have much lower performance when deployed. These were reflected in our generalization experiment results. We argue that others should routinely use these tests in publications. For industry, these tests could save time and money. Failing them indicates that the target objects' attributes are not being used by the systems undergoing analytical and clinical validation studies. Hence, as we show, relying only on conventional testing strategies with development data will not provide adequate assurances of generalization. Our sanity tests can be used with development data as long as ROIs are available or a background noise image can be generated.  While we focused on binary classification, the sanity tests apply to the multi-class classification and regression problems, with appropriate statistical analysis modifications. 

We did not attempt to use techniques to mitigate the impact of spurious correlations. These include adversarial regularization~\citep{ramakrishnan2018overcoming,zhang2018mitigating,ilyas2019adversarial}, model ensembling~\citep{cadene2019rubi,clark2019don}, invariant risk minimization~\citep{arjovsky2019invariant,choe2020empirical} and methods that encourage grounding on causal factors instead of spurious correlations~\citep{selvaraju2019taking,qi2020two,agarwal2020towards,castro2020causality}. However, as shown by Shrestha et al.~\citep{shrestha2020negative}, methods that were thought to overcome spurious correlations were behaving as regularizers instead of overcoming the issues that stemmed from the covariates. Our sanity tests could be used with these mitigation methods to measure their true impact, in that we would expect them to only be able to provide significant benefit when the target is present. 

There are some limitations to this study. As with most AI studies involving medical image analysis, we trained and tested with a small dataset. Results stemming from small-data may not always transfer to scenarios where larger datasets are used to train systems, but this is in part why sanity tests when using smaller datasets are critical since it is likely easier for spurious correlations to impact them. Our sanity tests help reveal when an AI model predicts the right answer for the wrong reasons and will therefore have a large gap between development and external generalization tests. A complementary approach uses visualization methods to understand if a system is not looking at the target to perform its classification. 

In conclusion, we demonstrated how our proposed sanity tests could identify spurious confounds early, using development data solely. While the methods are simple, we argue that sanity tests similar to these should be performed wherever possible, especially with smaller datasets, and if no external dataset is available. Otherwise, study results can be very misleading and fail to generalize on other datasets. In safety-critical AI domains, such as healthcare, sanity tests could prevent harm to patients, and they could better prepare novel medical AI systems for regulatory approval. We present a workflow and practical sanity tests that can reliably reveal error-prone systems before influencing real-world decision-making. 

\section*{Conflict of Interest Statement}
CK was employed at Paige, a commercial company, during the preparation of this manuscript.  This company played no role in the sponsorship, design, data collection and analysis, decision to publish, or preparation of the manuscript. The other authors declare that the research was conducted in the absence of any commercial or financial relationships that could be construed as a potential conflict of interest.

\section*{Author Contributions}
UM and CK conceived the study. UM implemented the algorithms and carried out the experiments. UM, CK, and RS wrote the paper. DB, LM, GC, and YE helped gather the data, provided advice, and reviewed the manuscript.

\section*{Funding}
This research was funded in part through the National Institutes of Health/National Cancer Institute Cancer Center support grant P30 CA008748 as well as by NSF award \#1909696. 


\section*{Data Availability Statement}
The datasets used to train for this study can be found in the TCIA Pancreas CT dataset, \url{https://wiki.cancerimagingarchive.net/display/Public/Pancreas-CT}, and in the Medical Image Segmentation Decathlon Pancreas Dataset, \url{http://medicaldecathlon.com/}.


\bibliographystyle{frontiersinSCNS_ENG_HUMS} 

\bibliography{References1}

\end{document}